\newif\ifOneCol
\OneColfalse 

\ifOneCol
    \documentclass[12pt,onecolumn,draftcls]{IEEEtran}
    \renewcommand{\baselinestretch}{1.7}
\usepackage{geometry}
\else
    \documentclass[11pt,final,twocolumn]{IEEEtran}
\usepackage{blindtext}
\usepackage{etoolbox}

\makeatletter
\def\do#1{\patchcmd{#1}{\thepage}{\null}{}{\GenericWarning{}{Could not patch \string#1}}}
\docsvlist{\@oddhead,\@evenhead,\ps@headings,\ps@IEEEtitlepagestyle,\ps@IEEEpeerreviewcoverpagestyle}
\makeatother

\usepackage{geometry}
    \renewcommand{\baselinestretch}{0.95}
\fi

\date{}
\usepackage{graphicx}
\usepackage{epsfig}
\usepackage{subfigure}
\usepackage{amssymb}
\usepackage{amsbsy}
\usepackage{amsmath}
\usepackage{steinmetz}
\usepackage[colorlinks=true, pdfborder={0 0 0}]{hyperref}
\usepackage{stmaryrd}
\usepackage{MnSymbol,  bbm} 
\usepackage{cite}
\usepackage{url}
\usepackage{svg}
\usepackage{amsfonts}
   
\usepackage{caption}

\usepackage{color}
\usepackage{float}
\usepackage{times,amsmath,epsfig}
\usepackage{xspace,latexsym,syntonly}
\usepackage{amssymb}
\usepackage{textcomp}

\usepackage{xcolor}
\usepackage{import}

\newtheorem{theorem}{Theorem}
\newtheorem{remark}{Remark}

\newcommand{\qed}{\hfill\blacksquare}
 
\DeclareMathOperator*{\argmin}{\arg\min} 

\usepackage{algorithm}
\usepackage{algpseudocode}

\usepackage{pgfplots, pgfplotstable}
\usepackage{tabularray}
\UseTblrLibrary{booktabs}

\usepackage{standalone}

\usepackage{tikz}
\pgfplotsset{compat=1.7}
\usepackage[utf8]{inputenc}

\begin{document}
\title{RRO: A Regularized Routing Optimization Algorithm for Enhanced Throughput and Low Latency with Efficient Complexity}
\author{David Zenati, Tzalik Maimon, Kobi Cohen \emph{(Senior Member, IEEE)}
\thanks{David Zenati and Kobi Cohen are with the School of Electrical and Computer Engineering, Ben-Gurion University of the Negev, Beer Sheva 8410501 Israel. Email: znatid@post.bgu.ac.il, yakovsec@bgu.ac.il.}
\thanks{Tzalik Maimon is with Ceragon Networks Ltd., Tel Aviv, Israel. Email: tzalikm@ceragon.com.}
\thanks{This work was supported by the Israel Ministry of Economy under the Magneton program.}
\thanks{This work has been submitted to the IEEE for possible publication. Copyright may be transferred without notice, after which this version may no longer be accessible.}
 }

\maketitle

\begin{abstract}
In the rapidly evolving landscape of wireless networks, achieving enhanced throughput with low latency for data transmission is crucial for future communication systems. While low complexity OSPF-type solutions have shown effectiveness in lightly-loaded networks, they often falter in the face of increasing congestion. Recent approaches have suggested utilizing backpressure and deep learning techniques for route optimization. However, these approaches face challenges due to their high implementation and computational complexity, surpassing the capabilities of networks with limited hardware devices.

A key challenge is developing algorithms that improve throughput and reduce latency while keeping complexity levels compatible with OSPF. In this collaborative research between Ben-Gurion University and Ceragon Networks Ltd., we address this challenge by developing a novel approach, dubbed Regularized Routing Optimization (RRO). The RRO algorithm offers both distributed and centralized implementations with low complexity, making it suitable for integration into 5G and beyond technologies, where no significant changes to the existing protocols are needed. It increases throughput while ensuring latency remains sufficiently low through regularized optimization. We analyze the computational complexity of RRO and prove that it converges with a level of complexity comparable to OSPF. Extensive simulation results across diverse network topologies demonstrate that RRO significantly outperforms existing methods.
\end{abstract}


\section{Introduction}
\label{sec:introduction}

The rapid evolution of communication network technology in 5G and beyond has led to a surge in demand for wireless communication services. Despite this progress, spectrum scarcity remains a significant constraint in meeting this growing demand. Hence, the development of efficient algorithms for data transmission in communication networks, which utilize available spectral resources and manage data transmissions effectively, has become a critical challenge in modern communication networks, such as 5G and beyond wireless and mobile networks \cite{papageorgiou2020advanced, cohen2020machine, iyer2022survey, patil2023case}, cognitive radio networks \cite{shah2020statistical, cohen2020machine, kaur2022comprehensive, gafni2022distributed, bavistale2023energy}, distributed \cite{sery2020analog, paul2021accelerated} and federated learning systems \cite{gafni2022federated, salgia2024communication}.

Recent applications in 5G and beyond technology require high demands on the network for higher throughput and low latency, especially in complex environments where multi-flow transmissions across network links interfere with each other, such as IoT, wireless sensor networks, drone networks, and autonomous vehicles. 

Low-complexity algorithms, like the shortest-path algorithms in OSPF-type solutions, are effective in lightly-loaded networks but struggle as congestion and mutual interference increase. Recent approaches advocate for route optimization using backpressure to alleviate congestion on heavily utilized preferred links. However, they face challenges such as frequent data exchanges between nodes and high storage and computational complexity as the network scales. Furthermore, deep learning methods have been proposed to boost throughput with low latency but require significant computational resources, exceeding the capabilities of networks with limited hardware devices. Therefore, a critical challenge in communication network research is to develop algorithms that achieve higher throughput with lower latency while keeping complexity levels compatible with OSPF.

In this collaborative research between Ben-Gurion University and Ceragon Networks Ltd., we address this challenge. In this context, this paper delves into the routing problem within communication networks with mutual interference environment. The network under consideration is represented as a directed connected graph \( \mathcal{G} = (\mathcal{V}, \mathcal{E}) \), where \( \mathcal{V} \) denotes the set of nodes, each corresponding to a user, and \( \mathcal{E} \) represents the communication links. Each directed link \( (v, u) \in \mathcal{E} \), connecting a transmitter \( v \) to a receiver \( u \) (where \( v, u \in \mathcal{V} \)), encapsulates the notion of neighboring nodes in this graph. A key characteristic of these links is their susceptibility to mutual interference, which arises when radio frequency signals overlap in the same spatial region, thereby affecting the link capacities adversely. We consider a network with \( N \) distinct flows, with each flow \( \phi_i \) originating from a source node \( s_i \) and directed towards a destination node \( d_i \), both of which are elements of \( \mathcal{V} \). The primary objective articulated in this research is the strategic allocation of routes for these multiple data flows. Such allocation must be optimized to enhance the network performance, reflecting a comprehensive approach to managing the inherent mutual interference. This interference, characteristic of concurrent transmissions along proximal routes, can significantly diminish the effective capacity of the involved links.

Hence, the challenge addressed here is to enhance the efficiency of spectral resource utilization by routing data flows through paths that enhance throughput and reduce delay, while simultaneously mitigating the detrimental effects of mutual interference between links across the network. This must be achieved using a low-complexity algorithm compatible with OSPF.

\subsection{Related Works}

Solving the shortest-path problem is a prevalent approach for data routing in contemporary communication networks \cite{heal2012ad, srikant2013communication, gong2016distributed}. This method aims to find the shortest path between a source and a destination node in a network, reducing the transmission delay and improving network performance. Shortest-path routing algorithms are widely used in various network protocols, such as OSPF and RIP. These protocols rely on the shortest-path algorithm to determine the most efficient routes for data transmission across the network. By selecting paths with the least number of hops or the lowest accumulated cost, these protocols aim to achieve efficient data routing and network operation. For example, the widely adopted OSPF routing protocol employs this principle to route data flows efficiently across the network, where the computation of the shortest paths is often facilitated by the Dijkstra algorithm \cite{dijkstra1959note, srikant2013communication, gong2016distributed}. More recent studies focus on online learning algorithms for shortest path computations under unknown link states that need to be inferred \cite{tehrani2013distributed, liu2012adaptive, amar2023online}. 

However, while shortest path-based routing is effective under various circumstances, it presents a significant drawback: It tends to increase congestion along the shortest routes, especially in networks experiencing high traffic loads. This congestion can, in turn, degrade overall network performance, as the most direct routes become highly congested with excessive data flow.

To address this limitation, backpressure routing emerges as an alternative routing aproach \cite{Backpressure_Survey}. Backpressure routing, derived from network utility maximization principles, dynamically adjusts data routes based on real-time congestion levels, contrasting sharply with traditional methods that focus on minimizing path cost \cite{Backpressure_first_work}. This strategy reduces congestion by maximizing differential queue backlog between nodes. It is known to achieve maximal throughput theoretically in a non-interference environment \cite{srikant2013communication}. Related studies on backpressure routing can be found in \cite{stolyar2005maximizing, eryilmaz2006joint, bui2009novel, ryu2010back, Athanasopoulou2013back, srikant2013communication, sinha2017optimal, joo2011performance, amar2023online} and references therein.  

Despite the efficiency of backpressure routing in terms of achievable throughput, it suffers from several main drawbacks. Firstly, backpressure routing tends to reduce congestion by transmitting packets through less congested nodes. While this approach is beneficial when congestion levels are high, it is inefficient when congestion is low. In such cases, packets travel along longer paths when shorter paths could have been used, leading to a significant increase in delay. Secondly, in an interference environment where links interfere with each other, transmitting data through longer paths increases the interference levels across the network, thereby reducing its overall capacity. Thirdly, from a practical implementation perspective, backpressure routing requires frequent message exchanges between neighboring nodes regarding their instantaneous queue states to make routing decisions. This necessity for constant updates for each transmission creates significant overhead for the communication protocol. These drawbacks have so far prevented the operational implementation of backpressure routing in real-world systems.

To address the limitation of increased delay in low congestion levels by backpressure routing, while incorporating the benefits of traditional shortest path routing, recent studies have suggested the development of hybrid routing algorithms. These approaches leverage the strengths of both methods, using shorter paths when congestion is low to avoid the delays characteristic of backpressure routing, and longer paths during high congestion to prevent the issues associated with shortest path routing \cite{ying2010combining, amar2023online}. This balance aims to optimize network performance across varying congestion levels, thereby increasing throughput while limiting path costs and adapting to dynamic network conditions. However, these methods still suffer from high implementation complexity, as they require frequent message exchanges between neighboring nodes regarding their instantaneous queue states to make backpressure-type routing decisions. Additionally, they necessitate storing a large number of buffers for each flow to optimize the route for each packet. Consequently, these methods are not practically appealing for a variety of network applications. 

In this paper, inspired by the hybrid approach as in \cite{ying2010combining, amar2023online}, we aim to strike a balance between increasing throughput and limiting path costs. However, a significant advantage of our approach is that we reduce congestion by seeking routes with high capacity, resulting in increased throughput without relying on backpressure. This avoids the drawbacks of backpressure-based routing discussed earlier. Additionally, our method maintains a complexity level compatible with OSPF.

Moreover, inspired by the recent success of deep learning and AI in solving complex decision-making tasks, deep learning methods have been proposed to solve related spectrum access \cite{wang2018deep, yu2019deep, naparstek2018deep, bokobza2023deep, cohen2024sinr}, scheduling \cite{budhiraja2020deep, Gahtan2022Deep, gahtan2023using, szostak2022decentralized, szostak2023deep}, and routing \cite{reis2019deep, reis2021r2l, DRL+GNN, paul2023multi, eyobu2023deep, yamin2024multi} problems. Specifically, recent studies developed deep learning based methods to compute flow routes in communication networks to increase throughput with low latency \cite{reis2019deep, reis2021r2l, DRL+GNN, paul2023multi, eyobu2023deep, yamin2024multi}. However, these methods demand high computational resources that exceed the capabilities of networks with limited hardware devices, making them unsuitable in many cases. Therefore, a key challenge in communication network research is to devise algorithms that achieve enhanced throughput with low latency while maintaining a complexity level compatible with OSPF.

Furthermore, combining low-complexity algorithms with deep-learning techniques could offer a versatile solution for future networks. This hybrid approach would allow for the flexibility to use deep learning when higher processing time is permissible and low-complexity algorithms when quick decisions are needed. Thus, these solutions do not compete but rather complement each other, offering a balanced and adaptive strategy to meet varying network demands.

\subsection{Main Results}
\label{sec:main_resutls}

In this paper, we concentrate on developing low-complexity algorithms that are compatible with OSPF. Our aim is to enhance network performance by achieving higher throughput and lower latency without increasing computational demands. These algorithms are designed to be efficient and scalable, ensuring they can be implemented in real-world networks with limited hardware resources. By aligning with the principles of OSPF, our approach maintains compatibility with existing network infrastructure, facilitating easy integration and adoption. Through this research, we seek to provide practical solutions that address the pressing challenges in modern communication networks.

Our motivation to balance enhanced throughput with low latency shares similarities with the approaches in \cite{ying2010combining, amar2023online}. However, our algorithmic solution is fundamentally different and has significant advancements compared to \cite{ying2010combining, amar2023online}. Firstly, it does not use backpressure for data transmission. Instead, it computes the balanced path to enhance capacity with low path cost at the source node and routes the packet through the computed path. This avoids the need for frequent message exchanges between neighboring nodes regarding their instantaneous queue states, as required by backpressure transmissions. By leveraging the advantages of 5G and beyond technologies, our method enables both distributed and centralized path computation, similar to current OSPF implementations. Consequently, no changes to the communication protocol are needed compared to current deployments. Secondly, the implementation complexity of our approach is much simpler as it does not require storing a large amount of packet buffers when making decisions, unlike \cite{ying2010combining, amar2023online}. Thirdly, the computational complexity of our approach is very low and is compatible with OSPF, as we will describe later. 

Specifically, in terms of algorithm development, we present a novel algorithm called Regularized Routing Optimization (RRO). The RRO algorithm offers both distributed and centralized implementations with low complexity, making it suitable for integration into 5G and beyond technologies with centralized node deployments. The RRO optimization process is implemented on a computation unit, either a source node of the flow in a distributed implementation or a centralized unit in a centralized implementation. This process utilizes the topology, link weights (such as link capacity), and an interference map, similar to the OSPF protocol used in 5G networks. The innovative design of the regularized optimization strikes a balance between the interference level over a route and the path length to yield the corresponding path allocation for each flow. It achieves this by searching for paths with high throughput while ensuring that latency remains sufficiently low.

Secondly, in terms of complexity analysis, we prove that RRO requires a low-complexity running time of 
\( O(N\cdot\vert{\mathcal{E}}\vert + N\cdot\vert{\mathcal{V}}\vert \log \vert{\mathcal{V}}\vert) \), where $\mathcal{E}$, and $\mathcal{V}$ denote the numbers of edges and nodes, respectively. This maintains the same complexity order as OSPF, as required. 

Finally, we extensively evaluated RRO's performance through simulations across various network topologies, including random deployments, NSFNET, and GEANT2 networks. The results consistently demonstrated that RRO significantly outperforms existing methods with compatible complexity. These simulations showcased RRO's ability to handle diverse and complex network conditions efficiently. The improvements were evident in terms of both throughput and latency, proving RRO's robustness and adaptability. Additionally, the algorithm's performance was stable across different scenarios, underscoring its potential for real-world applications in 5G and beyond technologies. This comprehensive evaluation highlights RRO as a superior routing solution that is very well-suited for modern network environments.

\section{Network Model and Problem Statement}
\label{sec:system}

We consider a routing problem in communication networks operating in an environment with mutual interferences between links. The communication network topology is conceptualized as a directed connected graph $\mathcal{G} = (\mathcal{V}, \mathcal{E})$, where each vertex $v \in \mathcal{V}$ represents a distinct node or user within the communication network. The set of edges (i.e. links) $\mathcal{E}$ of the graph corresponds to a set of communication channels between pairs of users. An edge $(i,j) \in \mathcal{E}$ exists between vertices $v_i\in\mathcal{V}$ and $v_j\in\mathcal{V}$ if and only if there is a communication link between user $i$ and user $j$. It is assumed that each communication channel is bidirectional, allowing for symmetrical data exchange between connected users. Links cause mutual interference due to the overlapping of their radio frequency signals in the same spatial region. 

Let $\mathcal{F}$ be a set of $N$ data flows, where each flow (say $f_n\in\mathcal{F}$) is indicated by source node $s_n\in\mathcal{V}$ and destination node $d_n\in\mathcal{V}$. A path (i.e., route) between $s_n$ and $d_n$ is denoted by $\pi=(s_n, v_1, v_2, ..., v_P, d_n)$, where $s_n, v_1, v_2, ..., v_P, d_n\in\mathcal{V}$, and $(s_n, v_1), (v_1, v_2), ..., (v_{P-1}, v_P), (v_P, d_n)\in\mathcal{E}$. A length of path $\pi$ is defined by the number of hops through the path: $\ell(\pi)=|\pi|-1$. Let $\pi_n$ be a selected path for flow $n$, $\Pi\triangleq(\pi_1, \pi_2, ..., \pi_N)$ be the selected path vector for all flows, and $\Pi_{-n}\triangleq(\pi_1, ..., \pi_{n-1}, \pi_{n+1},..., \pi_N)$ be the selected path vector for all flows excluding flow $n$. 

Let $R_n(\Pi)$ be the data-rate of flow $n$. Note that $R_n(\Pi)$ depends on the selected route of flow $n$ as well as the selected routes of other flows that interfere with flow $n$. We denote the set of flows that interfere with flow $n$ (i.e., interfering neighbors) by $\mathcal{N}_n$. All other flows $\mathcal{F}\setminus\mathcal{N}_n$ are transmitted through paths that cause interference below the noise floor (due to geographical distance, directional antennas, etc.). As a result, $R_n(\Pi)$ depends on the selected route of flow $n$, and the selected routes of flows $\mathcal{N}_n$, $R_n(\Pi)=R_n(\pi_n, \left\{\pi_i\right\}_{i\in\mathcal{N}_n})$. The cases of no-interference environment, and full-interference environment applies as special cases of our model, by setting $\mathcal{N}_n=\emptyset$, and $\mathcal{N}_n=\mathcal{F}\setminus f_n$, respectively. 

Let a congestion function $c(\Pi)$ be a monotonic decreasing function of the data rate \cite{srikant2013communication, cohen2015distributed}. Detailed computations of the data rate and congestion functions are discussed in Section \ref{sec:Alg}. 

The objective is to strike a balance between routing data through short paths to reduce the latency, and through low-congested paths to increase the throughput. Specifically, assume that $N-1$ paths were selected for flows $\mathcal{F}\setminus f_n$, given by path vector $\Pi_{-n}$. Then, the objective is to solve for flow $f_n$ the following optimization problem: 
\begin{equation}
\label{eq:opt}
    \pi_n = \argmin_{\pi} \; c(\pi, \Pi_{-n})+\lambda_n\cdot \ell(\pi),
\end{equation}
where $\lambda_n$ is a weight parameter. Increasing $\lambda_n$ tends to allocate a shorter path to flow $f_n$ at the expense of using more congested links, whereas decreasing $\lambda_n$ tends to assign a path with lower congestion to flow $f_n$, at the expense of using longer paths.

The solution is required to be of low complexity and compatible with OSPF. The solution is updated by each flow $f_n, n=1, ..., N$, periodically, given the current path allocation vector of other flows. In Section \ref{sec:Alg}, we develop both centralized and distributed implementations for solving \eqref{eq:opt}.

\remark{The optimization problem we address shares similarities with the one in \cite{ying2010combining, amar2023online} and subsequent studies, which combine shortest path and backpressure routing to mitigate congestion with low latency. However, there are significant differences between our optimization problem and algorithmic solutions compared to those in \cite{ying2010combining, amar2023online}. Specifically, the problem in \cite{ying2010combining, amar2023online} balances path length and queue size (a direct measure of congestion), leading to a complex backpressure-based routing algorithm. This approach suffers from high implementation complexity, as discussed in Section \ref{sec:introduction}. In contrast, our method reduces congestion by balancing path length with a congestion function that is monotonically decreasing with the data rate. This innovative approach allows us to develop a low-complexity end-to-end path solution at the source node without requiring backpressure transmissions. Additionally, the computational complexity of our method is very low and compatible with OSPF, as detailed in Section \ref{sec:Alg}.}

\section{The Regularized Routing Optimization (RRO) Algorithm}
\label{sec:Alg}

In this section, we introduce the RRO algorithm. This algorithm supports both distributed and centralized implementations with low complexity, making it well-suited for integration into 5G and next-generation technologies that use centralized node deployments. The RRO optimization process is executed on a computation unit, which can be either the source node in a distributed setup or a centralized unit in a centralized setup. This process leverages network topology, link weights (such as link capacity), and an interference map, similar to the OSPF protocol used in 5G networks. We start by presenting the distributed implementation of RRO, then discuss the necessary adjustments for the centralized implementation. The pseudocode of the algorithm is given in Algorithm \ref{alg:RRO}.

\subsection{Distributed RRO}
\label{ssec:distributed_RRO}
In the distributed implementation, each flow updates its route distributedly from time to time, as currently employed by OSPF mechanism, given the current path allocation vector of all other flows. Once a flow selects a path it broadcasts this information to all other source nodes in the network, to have each source node holds the updated path allocation vector of all other flows. 

We now describe the basic steps of RRO with respect to flow $f_n$, computed by its corresponding source node $s_n$. In the description below, we often omit the subscript $n$ when the notations become cumbersome, to simplify the presentation. Assume that $N-1$ paths were selected for flows $\mathcal{F}\setminus f_n$, given by path vector $\Pi_{-n}$.\vspace{0.2cm}\\
\noindent
\textbf{Initial computation step at source node $\boldsymbol{n}$:} Based on the interference map, and the 
current path allocation vector $\Pi_{-n}$ of all other flows, the source node $s_n$ computes the achievable rates of all links in the network: $\tilde{R}_l, \forall l\in\mathcal{E}$, assuming it selects link $l$ in its updated path. Specifically, the achievable rate over link $l$, $\tilde{R}_l(\Pi)$, is computed as follows: 
\begin{equation}
    \tilde{R}_{l}(\Pi)=
    B_l \cdot \log\left(1+\textrm{SINR}_l(\Pi)\right), 
    \label{eq:link_rate}
\end{equation}
where $B_l$ is the bandwidth of link $l$, and 
\begin{equation}
 \textrm{SINR}_l(\Pi)\triangleq\frac{P_l}{I_l(\Pi)
 +\sigma^2}   
\end{equation}
denotes the Signal to Interference plus Noise Ratio (SINR) at the reciever of link $l$. Here, $P_l$ is the received power of the signal, $\sigma^2$ is the power spectrum density (PSD) of the additive white Gaussian noise (AWGN), and 
\begin{equation}
I_l(\Pi) \triangleq \sum_{i\in\mathcal{N}_{l}}
I_{l,i}(\pi_i)
\end{equation}
is the cumulative interference power at the receiver of link $l$, aggregated across all links within its interference range $\mathcal{N}_l$. Here, $I_{l,i}$ is the interference power at the receiver of link $l$ caused by flow $i$, greater than zero if active by strategy $\pi_i$. The transmission power control mechanism considered in this paper is independent, where a predetermined transmission power is assigned to each link based on factors such as geographical distance and channel state. Consequently, the search space of the optimization problem is solely determined by the selected paths that affect the throughput.
\vspace{0.2cm}\\
\noindent
\textbf{Regularized Routing Optimization step at source node $\boldsymbol{n}$:} In this step, source node $s_n$ computes the updated path given the current path allocation vector $\Pi_{-n}$. 

Note that the rate of flow $f_n$ transmitted through path $\pi_n=(s_n, u_1, u_2, ..., d_n)$ is determined by the link with the lowest rate across the links in its path, say link $l^*$. The achievable rate of flow $n$ is thus given by:
\begin{equation}
    R_n(\Pi)=\tilde{R}_{l^*}(\Pi). 
    \label{eq:flow_rate}
\end{equation}

Therefore, the RRO algorithm aims at searching path with enhanced data rate, while limiting the path length to strike a trade-off between enhanced throughput and low latency. 

To allow this optimization procedure, we first apply linear transformation to transform the computed link rates into corresponding weights, where a link with higher capacity is assigned a lower weight, and vice versa. Let $\tilde{R}_{max} = \max_{e \in \mathcal{E}} \{\tilde{R}_{e}(\Pi)\}$, and  
$\tilde{R}_{min} = \min_{e \in \mathcal{E}} \{\tilde{R}_{e}(\Pi)\}$. Then, RRO computes:
\begin{equation}
\label{eq:inverse capacity}
w(e)=\frac{\tilde{R}_{max} - \tilde{R}_{e}(\Pi)}{\tilde{R}_{max} - \tilde{R}_{min}}, \;\forall e\in\mathcal{E},
\end{equation}
where the values $\tilde{R}_{e}(\Pi), \;\forall e\in\mathcal{E}$ were computed in the initial computation step. As a result, $w(e)$ is a mapping function from the link rate into $[0, 1]$, preserving the significance of order in the inverse direction.

\begin{algorithm}
\caption{The RRO Algorithm}
\label{alg:RRO}
\begin{algorithmic}[1]
\State \textbf{Inputs: }{$\mathcal{G} = (\mathcal{V}, \mathcal{E}), \{s_n, d_n, \lambda_n\}_{n=1}^{N}$}
\State \textbf{Output: }{$\Pi = \{\pi_n\}_{n=1}^{N}$}
\State   $\Pi \gets \varnothing$
\For{n = 1,...,N}
    \State $w_n(e) \gets Eq.\eqref{eq:inverse capacity}(\mathcal{G})\quad \forall e \in \mathcal{E}$
    \State $d_{n}(v) \gets \infty \quad \forall v \in \mathcal{V} \setminus\{s_n\}$
    \State $p_{n}(v) \gets \varnothing \quad \forall v \in \mathcal{V}$
    \State $h_{n}(v) \gets 0 \quad \forall v \in \mathcal{V}$
    \State $m_{n}(v) \gets 0 \quad \forall v \in \mathcal{V}$
    \State $d_{n}(s_{n}) \gets 0$
    \State $Q_{n} \gets \mathcal{V}$
    \While{$Q_{n} \text{ is not empty}$}
        \State $u \gets \text{vertex in } Q_{n} \text{ with\;} \min_{i\in Q_{n}} d_{n}(i)$
        \If{$u$ is $d_n$}
            \State {\textbf{break}}
        \EndIf
        \State $Q_{n} \gets Q_{n} \setminus \{u\}$
        \For{each neighbor $v \in Q$ of $u$}
        \State $\Tilde{m}_{n}(v) \leftarrow \max(m_{n}(u), w_{n}(u, v))$
        \State $\Tilde{h}_{n}(v) \leftarrow h_{n}(u) + \lambda_n$
        \State $\Tilde{d}_{n}(v) \leftarrow \Tilde{m}_{n}(v) + \Tilde{h}_{n}(v)$
            \If{$\tilde{d}_{n}(v) < d_{n}(v)$}
                \State $d_{n}(v) \gets \tilde{d}_{n}(v)$
                \State $p_{n}(v) \gets u$
                \State $m_{n}(v) \leftarrow \Tilde{m}_{n}(v)$
                \State $h_{n}(v) \leftarrow \Tilde{h}_{n}(v)$
            \EndIf
        \EndFor
    \EndWhile
    \State $\pi_n \gets Eq.\eqref{eq: path unfolding}(p_{n})$
    \State{Each user is notified about $\pi_n\in\Pi$\\
        \hspace{0.6cm}by $s_n$ (distributed implementation) or by \\
        \hspace{0.5cm} centralized unit (centralized implementation) }
\EndFor
    \State \Return $\Pi = \{\pi_n\}_{n=1}^{N}$
\end{algorithmic}
\end{algorithm}

Next, the congestion function is set to be:
\begin{equation}
\label{eq:congestion}
c(\pi,\Pi_{-n})=\max_{e\in\pi} \;w(e).
\end{equation}
Then, using this construction, the RRO algorithm can use $\lambda_n$ in \eqref{eq:opt} as a regularized parameter of flow $n$, to compute a low-complexity regularized optimization to update the selected path iteratively over the nodes. Increasing $\lambda_n$ increases the penalty of using long paths, with the price of increasing the congestion over the selected path. The procedure applies a Dijkstra-type update, were at each iteration, each node updates the index of its predecessor in the updated path through the network, and the updated regularized cost. The procedure starts with storing all nodes in a set $Q$, and initializing all regularized distances used in the computation to infinity, except the distance at the source node which is set to zero. 

First, the source node is removed from $Q$. Then, the regularized distance of all its neighbors is updated based on the following computation:
\begin{equation}
\label{eq:distance_source}
d(v) = w(s_n, v)  + \lambda_n \; \forall v: (s_n,v)\in\mathcal{E}, 
\end{equation}
and the index of the predecessor of $v$ is set to be $p(v)=s$. Then, RRO updates the hop counter (in units of $\lambda$) of node $u$ such that $h(v) = \lambda$ and set the maximal weight on the path from $s_n$ to $v$ to be $m(v) = w(s_n, v)$. In the next iteration, the node with the smallest regularized distance in $Q$ is selected (i.e., the neighbor of $s$ with the smallest regularized distance): $\min_{v:(s_n,v)\in\mathcal{E}} d(v)$. Assume node $u$ is selected for the next iteration. Then, it is removed from $Q$. Next, the regularized distance is computed with respect to all its neighbors:
\begin{equation}
\label{eq:distance_v}
\begin{array}{l}
\tilde{d}(v) = \max\left\{ m(u), w(u, v) \right\} + h(u)  + \lambda_n \;
\vspace{0.2cm}\\\hspace{5cm}
\forall v: (u, v)\in\mathcal{E}. 
\end{array}
\end{equation}
If $\tilde{d}(v)$ is smaller than the current value of $d(v)$, then RRO updates the value of $d(v)$ to be $d(v)=\tilde{d}(v)$, and the predecessor of $v$ is set to be $p(v)=u$. The procedure continues until convergence for flow $f_n$ (convergence to smallest regularized cost, and complexity analysis is provided in Section \ref{sec:XX_Alg_conv})

Then, the updated path is generated by the stored predecessors
\begin{equation}
\label{eq: path unfolding}
\pi_n=(s_n=p(u_1), u_1=p(u_2), ..., u_{\ell(\pi_n)-1}=p(d_n), d_n).
\end{equation}

Note that throughout the routine RRO searches for path with minimal regularized cost, consisting of the maximal value between $d(u)$ and $w(u,v)$ plus the regularization parameter $\lambda_n$. Decreasing $\lambda_n$ assigns more weight on reducing congestion to increase throughput, determined by the link with lowest data rate. while increasing $\lambda_n$ assigns more weight on minimizing the path length (i.e., tending to shortest path selection).\vspace{0.2cm}\\
\noindent
\textbf{Notification Step:} In the final step, the source node must inform all other source nodes of its selected route by broadcasting a control signal containing this information. This process does not significantly complicate the system compared to the current OSPF protocol, which already requires extensive message exchanges between nodes regarding network topologies and link weights before computing the shortest path. Therefore, it does not increase the implementation complexity compared to current deployments.

The procedure continues as each source node periodically updates its selected route. The pseudocode is given in Algorithm \ref{alg:RRO}

\subsection{Centralized RRO}
\label{ssec:distributed_RRO}

The centralized implementation of RRO follows similar steps to the distributed implementation, with the adjustments discussed in the following section. The key difference in the centralized implementation is that a central unit receives the input information regarding the network topology and interference map, and computes all routes for all nodes. It determines the desired path for each flow one by one and notifies each flow of its updated path. This approach aligns with 5G and beyond technology, which supports central unit deployments.\vspace{0.2cm}\\
\noindent
\textbf{Initial computation step at the centralized unit:} This step is similar to that in the distributed implementation, but it is performed at the centralized unit for flow $f_n$.\vspace{0.2cm}\\
\noindent
\textbf{Regularized Routing Optimization step at the centralized unit:} 
This step is similar to that in the distributed implementation, but it is performed at the centralized unit for flow $f_n$.\vspace{0.2cm}\\
\noindent
\textbf{Notification Step:} The centralized node must inform only source node $s_n$ of its selected route $\pi_n$ by
broadcasting a control signal containing this information.

The procedure continues as the centralized node repeat these steps for all flows to periodically update their selected routes. The pseudocode is given in Algorithm \ref{alg:RRO}.

\section{Convergence and Complexity Analysis of the RRO Algorithm}
\label{sec:XX_Alg_conv}
In this section, we analyze the convergence of RRO to the most efficient path in terms of regularized path cost, and the computational complexity of the algorithm. 
We denote the path that solves \eqref{eq:opt}, with congestion function $c(\pi,\Pi_{-n})$ given in \eqref{eq:congestion} as regularized most efficient path (RMEP). 

\begin{theorem}
\label{lemma:modified_dijkstra_correctness}
Consider the objective of solving \eqref{eq:opt} for flow $f_n$, with congestion function $c(\pi,\Pi_{-n})$ given in \eqref{eq:congestion}. Then, RRO solves \eqref{eq:opt}.
\end{theorem}

\emph{proof}
Throughout the proof, we refer to flow $n$. For simplicity, we often omit the subscript $n$. Let $\mathcal{V}{s} \subseteq \mathcal{V}$ represent a subset of vertices, where initially $\mathcal{V}{s} \leftarrow \varnothing$. This subset comprises vertices for which the true regularized most efficient paths from source $s$ have already been determined.
Let $d^{*}(u)$ represent the true RMEP of node $u$, signifying that this value remains invariant under any further relaxations. Correspondingly, the true RMEP from $s$ to node $u$, defined as an ordered sequence of vertices, is represented by $\pi^{*}_{u}$.

We shall prove the correctness of the RRO algorithm by induction, establishing the non-existence of any alternative RMEP.

For the base base case, initially, the RMEP to $s$ is set to 0, and to $\infty$ $ \forall v  \in \mathcal{V} \setminus \{s\} $. Clearly, \( d(s) = d^*(s) \) as the RMEP from \( s \) to itself is zero. 

For the inductive hypothesis, assume that after \( k \) iterations (where \( |\mathcal{V}_s| = k \)), the algorithm has correctly computed the RMEP for every node in \( \mathcal{V}_s \).

Next, in the inductive step, for the \((k+1)\)-th iteration, where \(|\mathcal{V}_s| = k + 1\), consider node \(u\) as having the minimum distance, \(d(u)\), among all nodes yet to be visited. We aim to demonstrate that the algorithm continues to compute the correct RMEP, i.e., that is, \(d(u) = d^*(u)\). We proceed by contradiction, assuming the existence of a hypothetical alternative path \( \tilde{\pi}_u \neq \pi_u \) with a cost \(\tilde{d}(u) \leq d(u)\). We analyze the implications of this assumption under two distinct scenarios that together cover all possibilities: (i) $\tilde{\pi}_{u}$ comprising nodes both within and outside $\mathcal{V}{s}$; (ii) $\tilde{\pi}_{u}$ consists solely of nodes from within \(\mathcal{V}_s\).

By examining these scenarios, we assert the non-existence of a valid \(\tilde{\pi}_u\) that would invalidate the algorithm's current RMEP calculation, thereby upholding the correctness of the computed paths at each iteration.

We start by contradicting case (i). At the $k + 1$ iteration, we add a node $u$ to $\mathcal{V}_s$. The following conditions are subsequently upheld: (a) $d(u) \leq d(\xi)$, $\forall \xi \in \mathcal{V} \setminus \mathcal{V}_{s}$ (according to the algorithm operation); (b) $d(p) = d(p)^{*}$, $\forall p \in \mathcal{V}_{s} \setminus \{u\}$ (according to induction hypothesis).

Let $(x_j, x_{j+1}) \in \mathcal{E}$ be the first edge taken from $\tilde{\pi}_{u}$, where $x_j \in \mathcal{V}_{s}$ and $x_{j+1} \in \mathcal{V} \setminus \mathcal{V}_{s}$,  Since we applied relaxation to $x_j$ when it was visited, we get:

\begin{equation}
\begin{array}{l}
\displaystyle
d(x_{j+1})
= \max \{m^{*}(x_{j}), w(x_{j}, x_{j+1})\}+ h^{*}(x_{j})  + \lambda,
\end{array}
\end{equation}
where:
\begin{equation}
    d^{*}(x_{j}) = m^{*}(x_{j}) + h^{*}(x_{j}).
\end{equation}
Moreover, since $d(x_{j+1})$ is only a subpath of $\tilde{\pi}_{u}$, we must have:

\begin{equation}
d(x_{j+1}) \leq \tilde{d}(u).
\end{equation}

Note that $\tilde{\pi}_{u}$ has a cost of $\tilde{d}(u) \leq d(u)$, and therefore $d(x_{j+1}) \leq \tilde{d}(u) \leq {d}(u)$. However, since $x_{j+1} \in \mathcal{V} \setminus \mathcal{V}_{s}$ we must have $d(u) \leq d(x_{j+1})$. Hence, a contradiction arises.

Next, we contradict case (ii). Let $\tilde{\pi}_{u}$ be 
$s \rightarrow x_1 \rightarrow \ldots \rightarrow x_j \rightarrow u$, where $s, x_1, \ldots, x_j \in \mathcal{V}_{s}$. According to the operations of the algorithm, we get:
\begin{equation}
\tilde{d}(u) =  \max \{m^{*}(x_{j}), w(x_{j}, u)\} + h^{*}(x_{j}) + \lambda  \leq d(u).
\end{equation}


Furthermore, we observe the following:
\begin{equation}
\begin{array}{l}
\displaystyle
d(x_j) = d^{*}(x_j) 
\vspace{0.2cm}\\
\displaystyle\hspace{0.5cm}
= m^{*}(x_{j}) + h^{*}(x_{j})
\vspace{0.2cm}\\
\displaystyle\hspace{0.5cm}
\leq \max \{m^{*}(x_{j}), w(x_{j}, u)\} + h^{*}(x_{j})
\vspace{0.2cm}\\
\displaystyle\hspace{0.5cm}
\leq \max \{m^{*}(x_{j}), w(x_{j}, u)\} + h^{*}(x_{j}) + \lambda
\vspace{0.2cm}\\
\displaystyle\hspace{0.5cm}
= \tilde{d}(u) \leq d(u). 
\end{array} 
\end{equation}

However, once $x_j$ is visited, it should have set $d(u)$ to at most  $ \max \{m^{*}(x_{j}), w(x_{j}, u)\} + h^{*}(x_{j}) + \lambda$. Hence,
\begin{equation}
\begin{array}{l}
\displaystyle
d(u) = \vspace{0.2cm}\\
\hspace{0.2cm}\min\{d(u), \max \{m^{*}(x_{j}), w(x_{j}, u)\} + h^{*}(x_{j}) + \lambda \} 
\vspace{0.2cm}\\
\displaystyle\hspace{0.5cm}
\leq \max \{m^{*}(x_{j}), w(x_{j}, u)\} + h^{*}(x_{j}) + \lambda
\vspace{0.2cm}\\
\displaystyle\hspace{0.5cm}
= \tilde{d}(u),
\end{array} 
\end{equation}
yielding a contradiction.

As a result, no such path \( \tilde{\pi}_u \) can exist, confirming the algorithm's correctness by induction.
$\qed$ \vspace{0.2cm}

As previously noted, our aim is to devise an algorithm for routing problems that is not only simple to implement but also maintains the same level of complexity as traditional, implemented in current deployments. Here, we establish that the complexity of our algorithm is commensurate with OSPF that implements the Dijkstra's algorithm when constructing the routing table.\vspace{0.2cm}

\begin{theorem} (Complexity Analysis of RRO Algorithm)
\label{lemma:Complexity Analysis of the RRO algorithm}
The RRO algorithm exhibits time complexity of $O(N\cdot|\mathcal{E}| + N\cdot|\mathcal{V}|\cdot \log(|\mathcal{V}|))$ and space complexity of $O(|\mathcal{E}| + N\cdot|\mathcal{V}|)$.
\end{theorem}
\emph{Proof:} 
To analyze the complexity of the RRO algorithm, we examine the computational cost associated with each step.

In the initial computation step, the source node calculates the achievable rates for all network links, resulting in time and space complexities of $O(|\mathcal{E}|)$.

Next, the initialization of predecessors and RMEP lists incurs time and space complexities of $O(|\mathcal{V}|)$.

Next, initializing the queue $Q$ requires $O(|\mathcal{V}|)$. Extracting the minimum from $Q$ traditionally requires $O(|\mathcal{V}|)$ per extraction, summing to $O(|\mathcal{V}^2|)$ for all vertices. However, utilizing a Fibonacci heap reduces this to $O(|\mathcal{V}|\cdot \log|\mathcal{V}|)$, as shown in \cite{dijksrta_heap}.

Finally, in the relaxation process, each neighbor's check and update operation requires complexity of $O(1)$. As a result, considering each edge once results in a total complexity of $O(|\mathcal{E}|)$.

Combining these findings, when allocating over $N$ data flows, the overall time complexity is $O(N\cdot(|\mathcal{E}| + |\mathcal{V}|\cdot \log|\mathcal{V}|))$, with the Q operations being the primary contributor. The total space complexity is $O(|\mathcal{E}| + N\cdot|\mathcal{V}|)$, accounted for by the initial computations and the storage needs of the queue, predecessors, and RMEP lists.
$\qed$ \vspace{0.2cm}

\section{Numerical Analysis}
\label{sec:analysis}

In this section, we present extensive experimental results to evaluate the performance of the proposed RRO algorithm under various wireless network topologies, namely NSFNET, GEANT2, and large-scale random network deployments. We conducted comparisons among well-known algorithms with compatible complexity: (i) \emph{Open Shortest Path First (OSPF)}: The OSPF protocol is extensively utilized for data routing and are implemented in numerous real-world systems\cite{srikant2013communication}.
In this context, we have implemented the well-established OSPF version that selects the route with the fewest number of hops between the source and the destination. This algorithm is widely employed in communication networks where minimizing latency is a critical objective\cite{becker2024maximizing}. 
(ii) \emph{Interference Minimization Algorithm (IMA):} Routing over paths with minimal interference is commonly used in environments where links experience mutual interference. In this context, we have implemented an interference minimization strategy that is compatible with OSPF's complexity, selecting paths that minimize absorbed interference to enhance throughput efficiency \cite{thelen2023survey}. (iii) \emph{Randomized Greedy Algorithm (RGA)}: RGA is a heuristic method that selects the best path allocation from $K=10$ independent trials of random choices from the action set. This algorithm is commonly used in machine learning research to assess a learning algorithm's capability to identify effective strategies and outperform exhaustive search within a constrained number of trials\cite{paul2023multi}. All selected algorithms exhibit the same complexity as our algorithm.

In the implementation of all tested algorithms, consistent knowledge of the wireless network state and the interference map was utilized to optimize the objective and determine the selected paths. For each simulation experiment, the results were averaged over $20$ randomly independent configurations, which included variations in flow demands and link capacities. In our analysis, we evaluated and presented the achievable data rate, averaged across data flows, the average delay experienced by packets as they traverse from the source node to the destination node, the log-rate of each flow as a fairness metric (i.e., proportional fairness), and the maximal queue length as a metric to evaluate the algorithms' efficacy in supporting data transmissions.

We categorized the simulated networks based on the node-to-link ratio. Specifically, networks where $N \leq 0.5\cdot|\mathcal{V}|$ are classified as lightly-loaded, those where $0.5\cdot|\mathcal{V}| < N \leq |\mathcal{V}|$ as moderately-loaded, and networks with $N > |\mathcal{V}|$ as heavily-loaded. This classification helped us systematically assess the network performance under varying load conditions. 

\subsection{Results for the NSFNET Network}

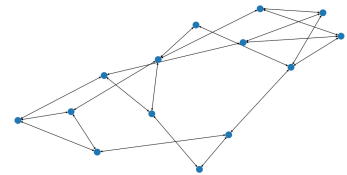
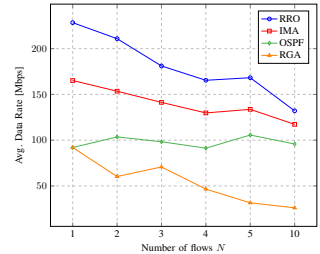
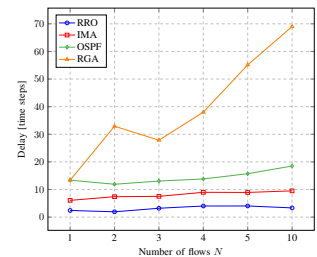
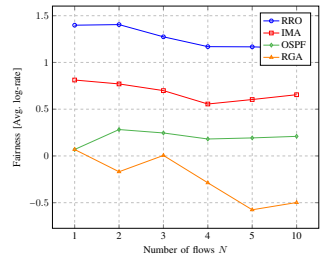
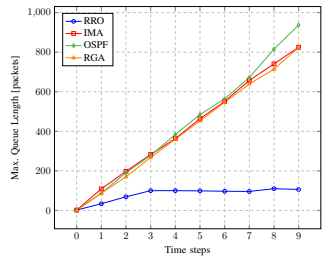
\begin{figure}[h!]
\centering
\subfigure[NSFNET]{
    \label{fig:NSFNET}
    \includegraphics[scale=0.2]{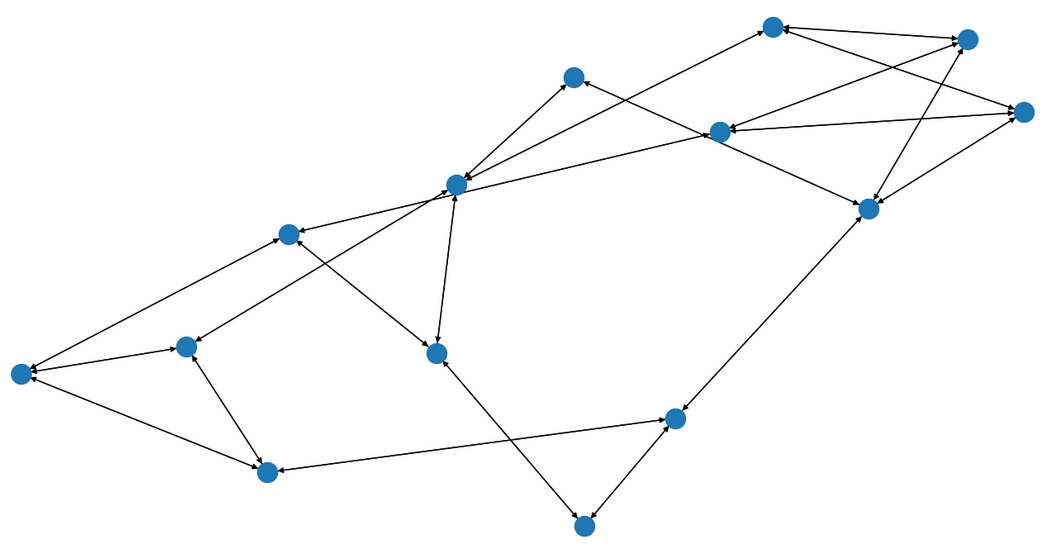}
}
\subfigure[Average Flows Rates]{
    \label{fig:NSFNET_rate}
    \scalebox{0.35}{
\pgfplotstableread[col sep=comma]{figures/nsfnet/avg_data_rate.csv}\data

\begin{tikzpicture}
\begin{axis}[
    name=plot1
    width=\textwidth,
    height=10cm,
    xtick=data,
    xticklabels from table={\data}{flows},
    legend pos=north east, 
    legend cell align={left},
    ylabel={Avg. Data Rate [Mbps]},
    xlabel={Number of flows $N$},
    grid=both,
    grid style=dashed,
    ]
    
    \addplot [mark=o, blue] table [x expr=\coordindex, y={RRO}]{\data};
    \addlegendentry{RRO}

    \addplot [mark=square, red] table [x expr=\coordindex, y={Dijkstra}]{\data};
    \addlegendentry{IMA}
    
    \addplot [mark=diamond, green!40!gray] table [x expr=\coordindex, y={OSPF}]{\data};
    \addlegendentry{OSPF}
    
    \addplot [mark=triangle, orange] table [x expr=\coordindex, y={RB}]{\data};
    \addlegendentry{RGA}
    
\end{axis}
\end{tikzpicture}}
}
\subfigure[Network Delay]{
    \label{fig:NSFNET_delay}
    \scalebox{0.35}{
\pgfplotstableread[col sep=comma]{figures/nsfnet/avg_trans_delay.csv}\data

\begin{tikzpicture}
\begin{axis}[
    name=plot1
    width=\textwidth,
    height=10cm,
    xtick=data,
    xticklabels from table={\data}{flows},
    legend pos=north west, 
    legend cell align={left},
    ylabel={Delay [time steps]},
    xlabel={Number of flows $N$},
    grid=both,
    grid style=dashed,
    ]
    
    \addplot [mark=o, blue] table [x expr=\coordindex, y={RRO}]{\data};
    \addlegendentry{RRO}

    \addplot [mark=square, red] table [x expr=\coordindex, y={Dijkstra}]{\data};
    \addlegendentry{IMA}
    
    \addplot [mark=diamond, green!40!gray] table [x expr=\coordindex, y={OSPF}]{\data};
    \addlegendentry{OSPF}
    
    \addplot [mark=triangle, orange] table [x expr=\coordindex, y={RB}]{\data};
    \addlegendentry{RGA}
    
\end{axis}
\end{tikzpicture}}
}

\subfigure[Fairness]{
    \label{fig:NSFNET_fairness}
    \scalebox{0.35}{
\pgfplotstableread[col sep=comma]{figures/nsfnet/avg_log_fair.csv}\data

\begin{tikzpicture}
\begin{axis}[
    name=plot1
    width=\textwidth,
    height=10cm,
    xtick=data,
    xticklabels from table={\data}{flows},
    legend pos=north east, 
    legend cell align={left},
    ylabel={Fairness [Avg. log-rate]},
    xlabel={Number of flows $N$},
    grid=both,
    grid style=dashed,
    ]
    
    \addplot [mark=o, blue] table [x expr=\coordindex, y={RRO}]{\data};
    \addlegendentry{RRO}

    \addplot [mark=square, red] table [x expr=\coordindex, y={Dijkstra}]{\data};
    \addlegendentry{IMA}
    
    \addplot [mark=diamond, green!40!gray] table [x expr=\coordindex, y={OSPF}]{\data};
    \addlegendentry{OSPF}
    
    \addplot [mark=triangle, orange] table [x expr=\coordindex, y={RB}]{\data};
    \addlegendentry{RGA}
    
\end{axis}
\end{tikzpicture}}
}
\subfigure[Max Queue Length $(N=10)$]{
    \label{fig:NSFNET_buffer}
    \scalebox{0.35}{
\pgfplotstableread[col sep=comma]{figures/nsfnet/max_buffer_nsfnet_N10.csv}\data

\begin{tikzpicture}
\begin{axis}[
    name=plot1
    width=\textwidth,
    height=10cm,
    legend pos=north west, 
    legend cell align={left},
    ylabel={Max. Queue Length [packets]},
    xlabel={Time steps},
    grid=both,
    grid style=dashed,
    ]
    
    \addplot [mark=o, blue] table [x expr=\coordindex, y={RRO}]{\data};
    \addlegendentry{RRO}

    \addplot [mark=square, red] table [x expr=\coordindex, y={Dijkstra}]{\data};
    \addlegendentry{IMA}
    
    \addplot [mark=diamond, green!40!gray] table [x expr=\coordindex, y={OSPF}]{\data};
    \addlegendentry{OSPF}
    
    \addplot [mark=triangle, orange] table [x expr=\coordindex, y={RB}]{\data};
    \addlegendentry{RGA}
    
\end{axis}
\end{tikzpicture}}
}

\caption{Performance Evaluation of the Algorithms in the NSFNET Network.}
\label{fig:nsfnet}
\end{figure}

We begin our evaluation by simulating the well-known NSFNET network topology \cite{NSFNET}, which comprises $|\mathcal{V}| = 14$ nodes and $|\mathcal{E}|=21$ edges, as illustrated in Fig. \ref{fig:NSFNET}. The results are presented in Figs. \ref{fig:NSFNET_rate}-\ref{fig:NSFNET_buffer}. It can be seen that RRO consistently outperforms all other evaluated algorithms, with an average data rate outperforming IMA by $10\%$ to $40\%$, as presented in Fig. \ref{fig:NSFNET_rate}. It is important to highlight that the superior data rate performance of the RRO algorithm does not compromise packet delay, as depicted in Fig. \ref{fig:NSFNET_delay}. Additionally, RRO demonstrates significantly higher fairness values compared to other algorithms, as illustrated in Fig. \ref{fig:NSFNET_fairness}. Finally, in Fig. \ref{fig:NSFNET_buffer} we present the maximum queue length for $N=10$, indicating that RRO's queues stabilize quickly, unlike other methods which diverge and fail to support high loads.

\subsection{Results for the GEANT2 Network}

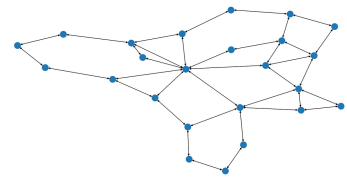
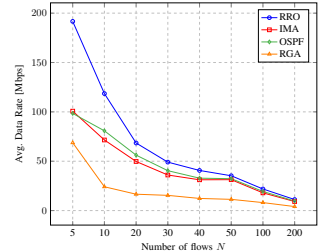
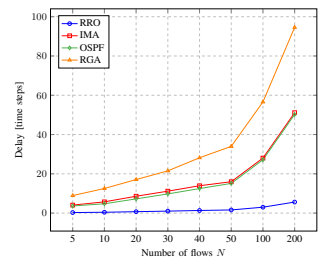
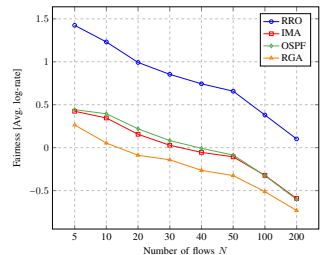
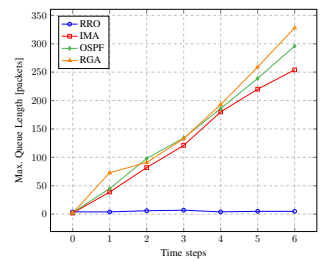
\begin{figure}[h!]
\centering
\subfigure[GEANT2]{
    \label{fig:GEANT2}
    \includegraphics[scale=0.2]{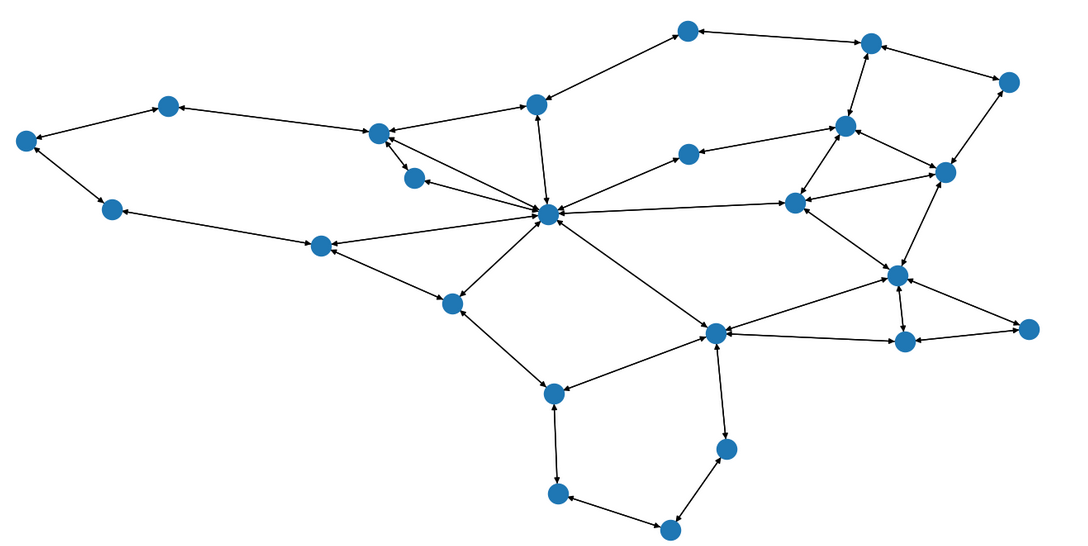}
}
\subfigure[Average Flows Rates]{
    \label{fig:GEANT2_rate}
    \scalebox{0.35}{
\pgfplotstableread[col sep=comma]{figures/geant2/avg_data_rate.csv}\data

\begin{tikzpicture}
\begin{axis}[
    name=plot1
    width=\textwidth,
    height=10cm,
    xtick=data,
    xticklabels from table={\data}{flows},
    legend pos=north east, 
    legend cell align={left},
    ylabel={Avg. Data Rate [Mbps]},
    xlabel={Number of flows $N$},
    grid=both,
    grid style=dashed,
    ]
    
    \addplot [mark=o, blue] table [x expr=\coordindex, y={RRO}]{\data};
    \addlegendentry{RRO}

    \addplot [mark=square, red] table [x expr=\coordindex, y={Dijkstra}]{\data};
    \addlegendentry{IMA}
    
    \addplot [mark=diamond, green!40!gray] table [x expr=\coordindex, y={OSPF}]{\data};
    \addlegendentry{OSPF}
    
    \addplot [mark=triangle, orange] table [x expr=\coordindex, y={RB}]{\data};
    \addlegendentry{RGA}
    
\end{axis}
\end{tikzpicture}}
}
\subfigure[Network Delay]{
    \label{fig:GEANT2_delay}
    \scalebox{0.35}{
\pgfplotstableread[col sep=comma]{figures/geant2/avg_trans_delay.csv}\data

\begin{tikzpicture}
\begin{axis}[
    name=plot1
    width=\textwidth,
    height=10cm,
    xtick=data,
    xticklabels from table={\data}{flows},
    legend pos=north west, 
    legend cell align={left},
    ylabel={Delay [time steps]},
    xlabel={Number of flows $N$},
    grid=both,
    grid style=dashed,
    ]
    
    \addplot [mark=o, blue] table [x expr=\coordindex, y={RRO}]{\data};
    \addlegendentry{RRO}

    \addplot [mark=square, red] table [x expr=\coordindex, y={Dijkstra}]{\data};
    \addlegendentry{IMA}
    
    \addplot [mark=diamond, green!40!gray] table [x expr=\coordindex, y={OSPF}]{\data};
    \addlegendentry{OSPF}
    
    \addplot [mark=triangle, orange] table [x expr=\coordindex, y={RB}]{\data};
    \addlegendentry{RGA}
    
\end{axis}
\end{tikzpicture}}
}

\subfigure[Fairness]{
    \label{fig:GEANT2_fairness}
    \scalebox{0.35}{
\pgfplotstableread[col sep=comma]{figures/geant2/avg_log_fair.csv}\data

\begin{tikzpicture}
\begin{axis}[
    name=plot1
    width=\textwidth,
    height=10cm,
    xtick=data,
    xticklabels from table={\data}{flows},
    legend pos=north east, 
    legend cell align={left},
    ylabel={Fairness [Avg. log-rate]},
    xlabel={Number of flows $N$},
    grid=both,
    grid style=dashed,
    ]
    
    \addplot [mark=o, blue] table [x expr=\coordindex, y={RRO}]{\data};
    \addlegendentry{RRO}

    \addplot [mark=square, red] table [x expr=\coordindex, y={Dijkstra}]{\data};
    \addlegendentry{IMA}
    
    \addplot [mark=diamond, green!40!gray] table [x expr=\coordindex, y={OSPF}]{\data};
    \addlegendentry{OSPF}
    
    \addplot [mark=triangle, orange] table [x expr=\coordindex, y={RB}]{\data};
    \addlegendentry{RGA}
    
\end{axis}
\end{tikzpicture}}
}
\subfigure[Max Queue Length $(N=30)$]{
    \label{fig:GEANT2_buffer}
    \scalebox{0.35}{
\pgfplotstableread[col sep=comma]{figures/geant2/max_buffer_geant2_N30.csv}\data

\begin{tikzpicture}
\begin{axis}[
    name=plot1
    width=\textwidth,
    height=10cm,
    legend pos=north west, 
    legend cell align={left},
    ylabel={Max. Queue Length [packets]},
    xlabel={Time steps},
    grid=both,
    grid style=dashed,
    ]
    
    \addplot [mark=o, blue] table [x expr=\coordindex, y={RRO}]{\data};
    \addlegendentry{RRO}

    \addplot [mark=square, red] table [x expr=\coordindex, y={Dijkstra}]{\data};
    \addlegendentry{IMA}
    
    \addplot [mark=diamond, green!40!gray] table [x expr=\coordindex, y={OSPF}]{\data};
    \addlegendentry{OSPF}
    
    \addplot [mark=triangle, orange] table [x expr=\coordindex, y={RB}]{\data};
    \addlegendentry{RGA}
    
\end{axis}
\end{tikzpicture}}
}

\caption{Performance evaluation of the algorithms in the GEANT2 network}
\label{fig:geant2}
\end{figure}

Next, we simulate the well-known GEANT2 network topology \cite{GEANT2}, which comprises $|\mathcal{V}|=24$ nodes and $|\mathcal{E}|=37$ edges, as illustrated in Fig \ref{fig:GEANT2}. The results are presented in Figs. \ref{fig:GEANT2_rate}-\ref{fig:GEANT2_buffer}. Once again, the RRO algorithm consistently surpasses all other evaluated algorithms, achieving an average data rate significant improvement, as presented in Fig. \ref{fig:GEANT2_rate}. We observe a significant improvement in lightly loaded networks, with the performance gap gradually narrowing. Ultimately, when the network is heavily loaded ($N \approx 4 \cdot |\mathcal{V}|$), the relative advantage of RRO compared to the other methods becomes less significant in terms of average data rate. In terms of average delay and fairness values, significant improvements were observed, where RRO continues to demonstrate substantial improvements even under extremely heavy loads settings, as illustrated in Fig. \ref{fig:GEANT2_delay} and in Fig. \ref{fig:GEANT2_fairness}, respectively.

Finally, Fig. \ref{fig:GEANT2_buffer} displays the maximum queue length observed in the temporal dynamics for a heavily loaded network with $N=30$ data flows
The findings reveal that the queue achieves stability quickly under RRO, unlike other methods which fail to support high loads.

\subsection{Results for Large-Scale Network deployments}

\begin{figure}[h!]
\centering
\subfigure[Random Topology $\mathcal{G} = (|\mathcal{V}| = 35, |\mathcal{E}| = 65)$]{
    \label{fig:random_topologyV35E65}
    \includegraphics[scale=0.2]{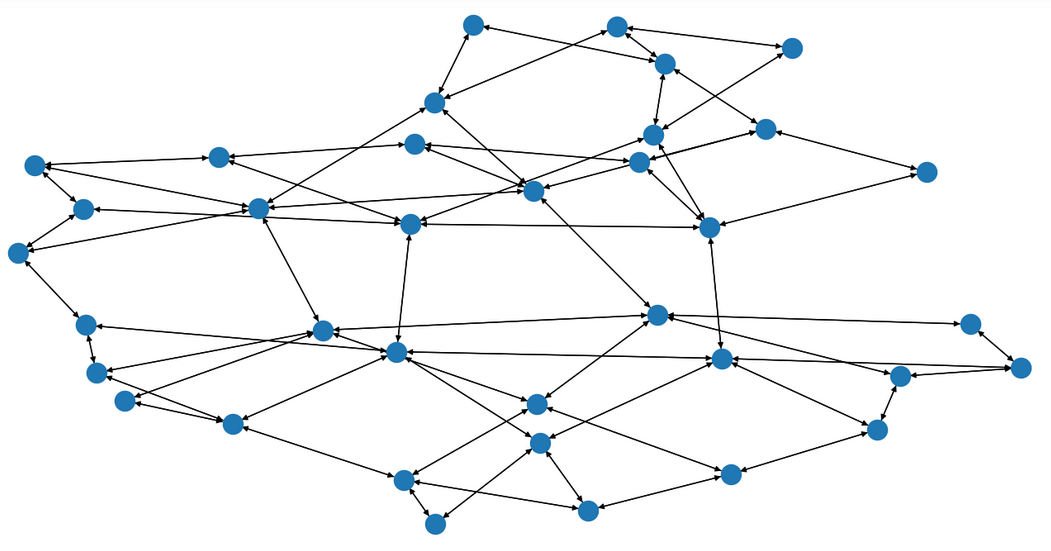}
}
\subfigure[Average Flows Rates]{
    \label{fig:random_topologyV35E65_rate}
    \scalebox{0.35}{
\pgfplotstableread[col sep=comma]{figures/V35E65/avg_data_rate.csv}\data

\begin{tikzpicture}
\begin{axis}[
    name=plot1
    width=\textwidth,
    height=10cm,
    xtick=data,
    xticklabels from table={\data}{flows},
    legend pos=north east, 
    legend cell align={left},
    ylabel={Avg. Data Rate [Mbps]},
    xlabel={Number of flows $N$},
    grid=both,
    grid style=dashed,
    ]
    
    \addplot [mark=o, blue] table [x expr=\coordindex, y={RRO}]{\data};
    \addlegendentry{RRO}

    \addplot [mark=square, red] table [x expr=\coordindex, y={Dijkstra}]{\data};
    \addlegendentry{IMA}
    
    \addplot [mark=diamond, green!40!gray] table [x expr=\coordindex, y={OSPF}]{\data};
    \addlegendentry{OSPF}
    
    \addplot [mark=triangle, orange] table [x expr=\coordindex, y={RB}]{\data};
    \addlegendentry{RGA}
    
\end{axis}
\end{tikzpicture}}
}
\subfigure[Network Delay]{
    \label{fig:random_topologyV35E65_delay}
    \scalebox{0.35}{
\pgfplotstableread[col sep=comma]{figures/V35E65/avg_trans_delay.csv}\data

\begin{tikzpicture}
\begin{axis}[
    name=plot1
    width=\textwidth,
    height=10cm,
    xtick=data,
    xticklabels from table={\data}{flows},
    legend pos=north west, 
    legend cell align={left},
    ylabel={Delay [time steps]},
    xlabel={Number of flows $N$},
    grid=both,
    grid style=dashed,
    ]
    
    \addplot [mark=o, blue] table [x expr=\coordindex, y={RRO}]{\data};
    \addlegendentry{RRO}

    \addplot [mark=square, red] table [x expr=\coordindex, y={Dijkstra}]{\data};
    \addlegendentry{IMA}
    
    \addplot [mark=diamond, green!40!gray] table [x expr=\coordindex, y={OSPF}]{\data};
    \addlegendentry{OSPF}
    
    \addplot [mark=triangle, orange] table [x expr=\coordindex, y={RB}]{\data};
    \addlegendentry{RGA}
    
\end{axis}
\end{tikzpicture}}
}

\subfigure[Fairness]{
    \label{fig:random_topologyV35E65_fairness}
    \scalebox{0.35}{
\pgfplotstableread[col sep=comma]{figures/V35E65/avg_log_fair.csv}\data

\begin{tikzpicture}
\begin{axis}[
    name=plot1
    width=\textwidth,
    height=10cm,
    xtick=data,
    xticklabels from table={\data}{flows},
    legend pos=north east, 
    legend cell align={left},
    ylabel={Fairness [Avg. log-rate]},
    xlabel={Number of flows $N$},
    grid=both,
    grid style=dashed,
    ]
    
    \addplot [mark=o, blue] table [x expr=\coordindex, y={RRO}]{\data};
    \addlegendentry{RRO}

    \addplot [mark=square, red] table [x expr=\coordindex, y={Dijkstra}]{\data};
    \addlegendentry{IMA}
    
    \addplot [mark=diamond, green!40!gray] table [x expr=\coordindex, y={OSPF}]{\data};
    \addlegendentry{OSPF}
    
    \addplot [mark=triangle, orange] table [x expr=\coordindex, y={RB}]{\data};
    \addlegendentry{RGA}
    
\end{axis}
\end{tikzpicture}}
}
\subfigure[Max Queue Length $(N=40)$]{
    \label{fig:random_topologyV35E65_buffer}
    \scalebox{0.35}{
\pgfplotstableread[col sep=comma]{figures/V35E65/max_buffer_V35E65_N40.csv}\data

\begin{tikzpicture}
\begin{axis}[
    name=plot1
    width=\textwidth,
    height=10cm,
    legend pos=north west, 
    legend cell align={left},
    ylabel={Max. Queue Length [packets]},
    xlabel={Time steps},
    grid=both,
    grid style=dashed,
    ]
    
    \addplot [mark=o, blue] table [x expr=\coordindex, y={RRO}]{\data};
    \addlegendentry{RRO}

    \addplot [mark=square, red] table [x expr=\coordindex, y={Dijkstra}]{\data};
    \addlegendentry{IMA}
    
    \addplot [mark=diamond, green!40!gray] table [x expr=\coordindex, y={OSPF}]{\data};
    \addlegendentry{OSPF}
    
    \addplot [mark=triangle, orange] table [x expr=\coordindex, y={RB}]{\data};
    \addlegendentry{RGA}
    
\end{axis}
\end{tikzpicture}}
}

\caption{Performance evaluation of the algorithms in large-scale random network.}
\label{fig:performance_random_V35E65}
\end{figure}
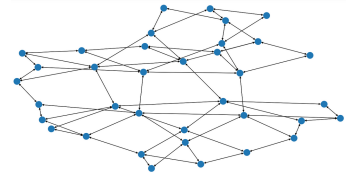
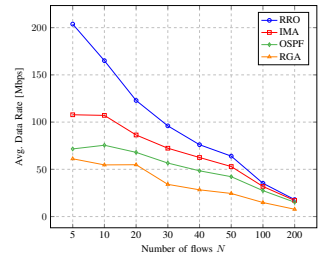
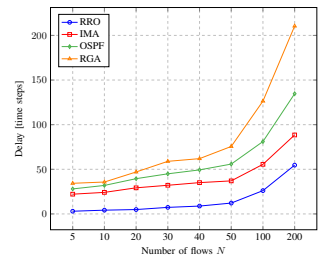
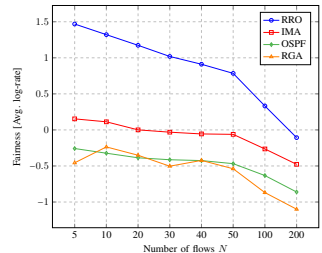
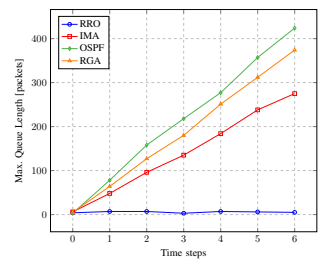

Finally, we conducted simulations on random large-scale network deployments. First, we present the results on a random medium-size network configuration (presented in Fig. \ref{fig:random_topologyV35E65}) which comprises $V=35$ nodes and $E=65$ random edges between nodes. 
The results are presented in Figs. \ref{fig:random_topologyV35E65_rate}-\ref{fig:random_topologyV35E65_buffer}.
Across all scenarios, RRO consistently outperforms all other algorithms, particularly in lightly and moderately loaded networks. In terms of average flow data rates (Fig. \ref{fig:random_topologyV35E65_rate}), RRO surpasses the second-best performer, IMA, by margins ranging from  $90\%$ (where $N=5$) to $5\%$ (where $N=200$). Furthermore, RRO demonstrates superior performance in reducing flow delays (Fig. \ref{fig:random_topologyV35E65_delay}) and enhancing fairness (Fig. \ref{fig:random_topologyV35E65_fairness}), even under extreme network loads. We also evaluated the queue length for each algorithm where $N=40 > |\mathcal{V}|$ (Fig. \ref{fig:random_topologyY_buffer}). The findings indicate that the maximum queue length stabilizes quickly by RRO, whereas other algorithms fail to support high loads. 

To further substantiate the robustness of the RRO algorithm in accommodating diverse network configurations characterized by varying numbers of flow demands $N$, nodes $|\mathcal{V}|$, and edges $|\mathcal{V}|$, and to validate its applicability in 5G topologies, which typically comprise dense networks with numerous edges, additional experiments were conducted using randomized settings. The outcomes of these experiments are detailed in Table \ref{tab:rand_performances}. The simulation results demonstrate that our method excels in both heavily-loaded and moderately-loaded dense networks (lines \hyperlink{r1}{1} and \hyperlink{r2}{2}, respectively), where \(|\mathcal{E}| > N > |\mathcal{V}|\). Continuing in line \hyperlink{r3}{3}, a scenario is presented where \(|\mathcal{E}| \gg N = |\mathcal{V}|\), indicative of a highly connected dense network. Additionally, a scenario with a moderate degree of connectivity under heavy load, where \(|\mathcal{E}| = N > |\mathcal{V}|\), is examined in line \hyperlink{r4}{4}. The outcomes, detailed in Table \ref{tab:rand_performances}, confirm that RRO significantly surpasses all other algorithms in terms of average flow rate, packet delay and fairness across all tested network configurations.

These results underscore the notable superiority of RRO over alternative algorithms in all tested scenarios, showcasing its consistent performance advantage. The robustness and efficacy of RRO are evident, presenting a compelling case for its adoption in diverse network environments.

\begin{table*}[!htpb]

	\caption{Algorithm comparison for various network configurations. 
}
	\label{tab:rand_performances}
	\begin{center}
		\scalebox{0.9}{
			\begin{tabular}{c c c || c c c c | c c c c | c c c c}
                \toprule
                & & & \multicolumn{4}{c|}{Average Flow Rates [Mbps] $\uparrow$} & \multicolumn{4}{c|}{Average Delay [time-steps] $\downarrow$} & \multicolumn{4}{c}{Fairness $\uparrow$}\\
				\midrule 
				$N$ & $|\mathcal{V}|$ & $|\mathcal{E}|$ & RGA & OSPF & IMA & \textbf{RRO} & RGA & OSPF & IMA & \textbf{RRO}
                & RGA & OSPF & IMA & \textbf{RRO}\\ \hline
    
				\hypertarget{r1}{}100 & 70 & 130 &
                3.839 & 5.310 & 6.679 &  \textbf{9.389}
                & 108.608 & 74.396 & 43.543 &  \textbf{18.395}
                & -0.728 & -0.557 & -0.164 &  \textbf{0.565}\\ 

                \hypertarget{r2}{}50 & 60 & 120 &
                5.425 & 6.530 & 8.468 &  \textbf{13.582}
                & 91.985 & 79.509 & 54.112 &  \textbf{18.664}
                & -0.655 & -0.644 & -0.265 &  \textbf{0.730}\\ 

                \hypertarget{r3}{}40 & 40 & 100 &
                9.757 & 11.287 & 18.972 &  \textbf{23.790}
                & 47.586 & 39.699 & 18.073 &  \textbf{2.75}
                & -0.345 & -0.340 & 0.455 &  \textbf{1.260}\\ 

                \hypertarget{r4}{}50 & 30 & 50 &
                3.027 & 6.113 & 6.799 &  \textbf{7.689}
                & 127.832 & 92.847 & 72.830 &  \textbf{43.028}
                & -1.011 & -0.783 & -0.556 &  \textbf{0.092}\\ 

                \hypertarget{r5}{}30 & 25 & 40 &
                6.413 & 10.929 & 12.432 &  \textbf{14.804}
                & 78.615 & 39.376 & 34.426 &  \textbf{10.826}
                & -0.604 & -0.263 & -0.017 &  \textbf{0.825}\\

            \bottomrule
			\end{tabular}
		}
	\end{center}
\end{table*}

\section{Conclusion}
\label{sec:conclusion}
Achieving enhanced throughput with low latency for data transmission is crucial for future communication systems. Low-complexity OSPF-type solutions are effective in lightly-loaded networks but struggle with increased congestion. We addressed this challenge by developing a novel approach: Low-complexity Regularized Routing Optimization (RRO) in communication networks. The RRO algorithm offered both distributed and centralized implementations with low complexity, making it suitable for integration into 5G and beyond technologies. It increased throughput while ensuring latency remained sufficiently low through regularized optimization. We analyzed the convergence complexity of RRO and proved that it converged with a complexity order comparable to OSPF. Our simulation results demonstrated significant performance improvements over existing methods.

These findings position RRO as a highly effective solution for routing data in modern communication networks, offering both enhanced efficiency and low complexity.

\bibliographystyle{IEEEtran}

\end{document}